\documentclass{article}
\usepackage{spconf,amsmath,graphicx}
\usepackage{txfonts}
\usepackage{enumitem}
\usepackage{color,xcolor}
\usepackage{colortbl}
\usepackage{threeparttable}
\usepackage{subfigure}
\usepackage[colorlinks=true]{hyperref}
\usepackage{array}
\newcommand{\PreserveBackslash}[1]{\let\temp=\\#1\let\\=\temp}
\newcolumntype{C}[1]{>{\PreserveBackslash\centering}p{#1}}
\newcolumntype{R}[1]{>{\PreserveBackslash\raggedleft}p{#1}}
\newcolumntype{L}[1]{>{\PreserveBackslash\raggedright}p{#1}}

\title{Unsupervised learning based end-to-end delayless generative fixed-filter active noise control}

\name{Zhengding Luo, Dongyuan Shi, Xiaoyi Shen, Woon-Seng Gan
\thanks{The code will be available at \href{https://github.com/Luo-Zhengding/Unsupervised-GFANC}{https://github.com/Luo-Zhengding/Unsupervised-GFANC}. This research is supported by the Singapore Ministry of Education, Academic Research Fund Tier 2, under research grant MOE-T2EP20221-0014.}}
\address{School of Electrical \& Electronic Engineering, Nanyang Technological University, Singapore.\\
Emails: LUOZ0021@e.ntu.edu.sg; dongyuan.shi@ntu.edu.sg; ewsgan@ntu.edu.sg}

\begin{document}
%\ninept % 整体压缩行距
\maketitle

\begin{abstract}
Delayless noise control is achieved by our earlier generative fixed-filter active noise control (GFANC) framework through efficient coordination between the co-processor and real-time controller. However, the one-dimensional convolutional neural network (1D CNN) in the co-processor requires initial training using labelled noise datasets. Labelling noise data can be resource-intensive and may introduce some biases. In this paper, we propose an unsupervised-GFANC approach to simplify the 1D CNN training process and enhance its practicality. During training, the co-processor and real-time controller are integrated into an end-to-end differentiable ANC system. This enables us to use the accumulated squared error signal as the loss for training the 1D CNN. With this unsupervised learning paradigm, the unsupervised-GFANC method not only omits the labelling process but also exhibits better noise reduction performance compared to the supervised GFANC method in real noise experiments.

\end{abstract}

\vspace*{-0.3cm}

\section{Introduction}\vspace*{-0.2cm}
Increasing urban noise pollution has detrimental effects on environmental quality and human health. As a potential solution, active noise control (ANC) technology can generate an anti-noise that is phase-inverted and has the same amplitude as the undesired noise \cite{1,2}. The combination of the anti-noise with the unwanted sound reduces the overall sound level through destructive interference \cite{3,4}. Compared to passive noise control strategies, such as barriers and sound absorbers, ANC techniques exhibit more effectiveness in size, cost, and efficiency in dealing with low-frequency noises \cite{5,6}. Thus, ANC has been widely used in many fields including headphones, ventilation ducts, automobiles, etc \cite{26,27}.

Traditional ANC systems often adopt adaptive algorithms to continuously update control filter coefficients for minimizing error signals \cite{9,23,24}. Especially, the filtered-X least mean square (FxLMS) algorithm and its variants are commonly used, as they can address the delays associated with the secondary path and achieve high computational efficiency \cite{13,14}. Nevertheless, the adaptive ANC algorithms are limited by slow response time and the potential risk of divergence, which could negatively impact the overall noise reduction performance \cite{12,11}. To maintain system stability and alleviate the need for complex adaptation procedures, fixed-filter ANC algorithms are extensively employed in practice.

However, the fixed-filter ANC algorithms are optimized for specific noise types, resulting in mediocre performance when controlling other types of noise \cite{15}. To generate appropriate control filters for different primary noises, a generative fixed-filter active noise control (GFANC) approach was proposed in our previous work \cite{17,28,29}. Specifically, the one-dimensional convolutional neural network (1D CNN) in the co-processor can predict the combination weights of sub control filters given the incoming noise. A new control filter is then generated by combining sub control filters and used for real-time noise control.

While the GFANC method exhibits effective noise reduction performance, it necessitates the supervised training of the 1D CNN using labelled noise datasets. Labelling noise data typically requires a significant amount of resources and has the potential to introduce errors \cite{16,20}. As an alternative approach, unsupervised learning can leverage unlabeled data to train the network, reducing annotation costs and mitigating certain biases linked to manual labelling \cite{18,19}. Furthermore, given the prevalence of unlabeled noise data in the ANC field, the incorporation of unsupervised learning can greatly enhance the practicality of ANC techniques \cite{21,25}.

However, unsupervised learning is still a nascent research area in ANC techniques. This paper proposes an unsupervised-GFANC method that omits the labelling process. In this method, the co-processor and real-time controller are integrated into an end-to-end derivable ANC system during training. By utilizing the accumulated squared error signal as loss for training the 1D CNN, its parameters can be automatically updated through a derivative mechanism. Experimental results indicate that the unsupervised-GFANC method outperforms the supervised GFANC method in attenuating real noises due to eliminating annotation errors.

\vspace*{-0.2cm}
\section{The Unsupervised-GFANC Approach}
\begin{figure}[tp]
\centering
\centerline{\includegraphics[width=0.9\linewidth]{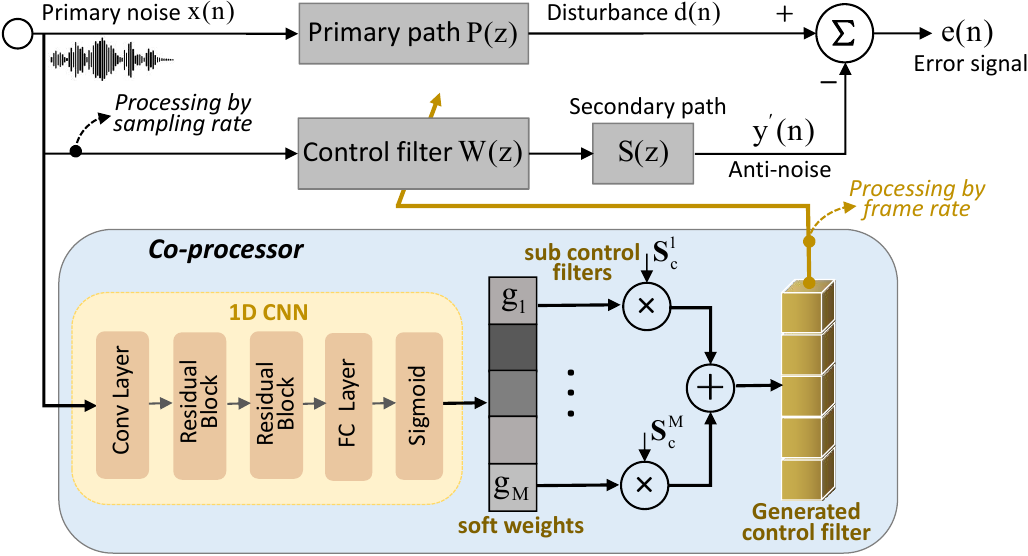}}\vspace*{-0.1cm}
\caption{Framework of the unsupervised-GFANC method. After unsupervised training, the 1D CNN collaborates with the real-time controller for delayless noise control.}
\label{Fig 1}\vspace*{-0.3cm}
\end{figure}

\begin{figure}[tp]
\centering
\centerline{\includegraphics[width=0.9\linewidth]{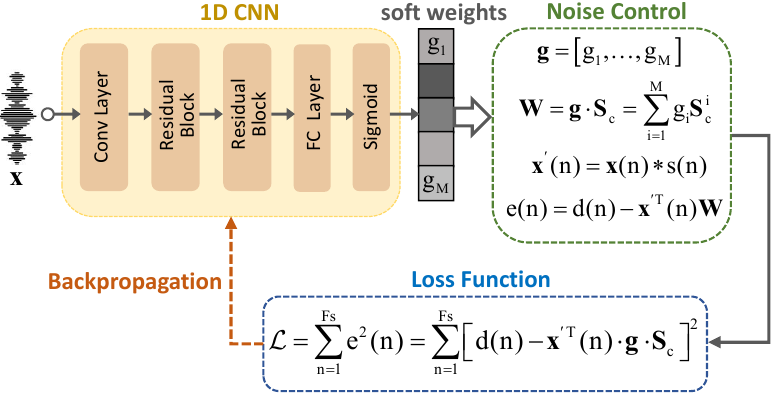}}\vspace*{-0.1cm}
\caption{Block diagram of training the 1D CNN in the unsupervised-GFANC approach, which is an end-to-end differentiable ANC system. Given each noise frame, the accumulated squared error signal is used as the training loss.}
\label{Fig 2}\vspace*{-0.2cm}
\end{figure}

The framework of the unsupervised-GFANC approach is illustrated in Fig.~\ref{Fig 1}. During training the 1D CNN, the co-processor and real-time controller are integrated into an end-to-end differentiable ANC system, as shown in Fig.~\ref{Fig 2}. Given each noise frame, we employ the accumulated squared error signal from the end-to-end system as the training loss. With the backpropagation of derivatives, the 1D CNN's parameters can be automatically updated without noise labels. After unsupervised training, the co-processor operates at the frame rate while the real-time controller operates at the sampling rate. The efficient coordination between the co-processor and real-time controller enables delayless noise control in Fig.~\ref{Fig 1}.

\vspace*{-0.3cm}
\subsection{The 1D CNN}
The co-processor in unsupervised-GFANC mainly performs the weighted sum of sub control filters to generate new control filters. Sub control filters are obtained by decomposing a pre-trained broadband control filter \cite{17}. To adaptively combine sub control filters, a lightweight 1D CNN is constructed to output the combination weights of sub control filters for various noises. In this task, we use soft weights to combine sub control filters, which implies that the weights are numerical values ranging from 0 to 1. Hence, using the 1D CNN to produce the soft weights belongs to a regression task. The next section will explain how to use an unsupervised paradigm to train the 1D CNN for the regression task.

\vspace*{-0.2cm}
\subsection{Unsupervised Training}
As introduced in \cite{29}, a practical filter decomposition technique based on the theory of filter perfect-reconstruction is used to obtain sub control filters. The sub control filter bank $\mathbf{S}_{c}$ composed of $M$ sub control filters serves as the orthogonal base for generating control filters and is given by
\begin{equation}
\mathbf{S}_{c} = \left[\mathbf{S}^{1}_c, \ldots, \mathbf{S}^{m}_c, \ldots, \mathbf{S}^{M}_c \right].
\end{equation}
Given each noise frame $\mathbf{x}$, the 1D CNN model can output the corresponding weight vector $\mathbf{g}$ in the co-processor:
\begin{equation}
\begin{aligned}
\mathbf{g} &=\operatorname{CNN}\left(\mathbf{x}\right),\\
\mathbf{g} &= \left[g_1, \ldots, g_m, \ldots, g_M \right],
\end{aligned}
\end{equation}
where $\operatorname{CNN}(\cdot)$ denotes the operation of the 1D CNN model. A new control filter $\mathbf{W}$ is generated by computing the inner product between the weight vector $\mathbf{g}$ and the sub control filter bank $\mathbf{S}_{c}$ as
\begin{equation}
\mathbf{W} = \mathbf{g} \cdot \mathbf{S}_{c} = \sum_{i=1}^{M} g_i \mathbf{S}^{i}_c.
\end{equation}

The error signal following noise control is calculated using the generated control filter $\mathbf{W}$ as
\begin{equation}
\begin{cases}
e(n)&= d(n)-\mathbf{x}^{\prime\mathrm{T}}(n)\mathbf{W} \\
\mathbf{x}^\prime(n)&= \mathbf{x}(n)*s(n),
\end{cases}
\end{equation}
where $s(n)$ and $*$ denote the impulse response of the secondary path and the linear convolution, respectively. The filtered reference signal $\mathbf{x}^\prime(n)$ refers to the reference signal being filtered by the estimated secondary path.

To obtain the best 1D CNN that assists the unsupervised-GFANC system in achieving optimal noise control, we defined its loss function as
\begin{equation}
\mathcal{L} =\sum_{n=1}^{F_s} e^2(n) = \sum_{n=1}^{F_s} \left[d(n)-\mathbf{x^{\prime\mathrm{T}}}(n) \cdot \mathbf{g} \cdot \mathbf{S}_{c} \right]^2.
\end{equation}
where $F_s$ represents the number of samples in each noise frame. It is found that the loss function $\mathcal{L}$ is a differentiable function with respect to the weight vector $\mathbf{g}$.

Table \ref{Table 1} outlines the pseudo-code for calculating the training loss. Each training iteration collectively processes a batch of noise frames to achieve faster convergence and enhanced generalization. The whole process of loss calculation is derivable in Table \ref{Table 1}. In this way, the co-processor and real-time controller are integrated into an end-to-end derivable ANC system. Hence, the 1D CNN's parameters can be automatically updated in the backpropagation process of derivatives, achieving unsupervised learning.

\vspace*{-0.5cm}
\subsection{Comparing Unsupervised and Supervised Training}
\begin{figure}[tp]
\centering
\includegraphics[width=0.8\linewidth]{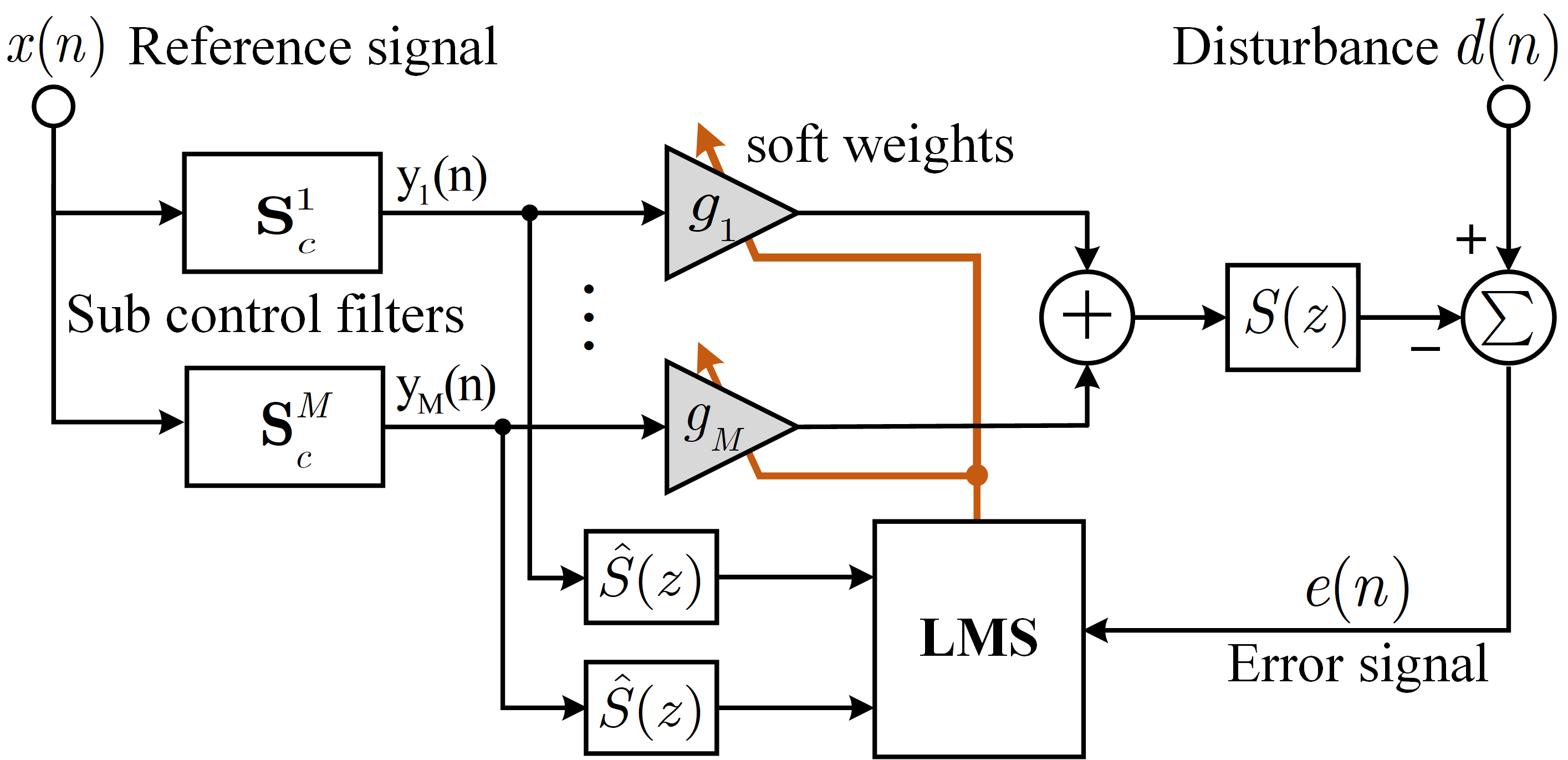}\vspace*{-0.2cm}
\caption{Block diagram of the labelling mechanism in the previous supervised GFANC method.}
\label{Fig 3}\vspace*{-0.4cm}
\end{figure}

The previous supervised GFANC method~\cite{17} employed an adaptive labelling mechanism, as depicted in Fig.~\ref{Fig 3}, to assign labels to the noise instances in the training dataset. Following the labelling process, the noise instances and their respective soft weights are utilized to train the 1D CNN. To obtain the labels of soft weights for combining sub control filters, the least mean square (LMS) algorithm is used to update the soft weights. However, since the adaptive labelling mechanism is based on the LMS algorithm, it is susceptible to inappropriate step size and slow convergence. As a result, the labelling process may introduce additional biases and is time-consuming.

\begin{table}[tp]
\centering
\caption{Pseudo-code of computing the error signal as the training loss in the unsupervised-GFANC method}\vspace*{-0.1cm}
\begin{tabular}{l}
\multicolumn{1}{l}{\rule{8.5cm}{0.2pt}}\\
\textbf{Input:} $B$ refers to the batch size during training the 1D CNN. \\ $\mathbf{W}$: the generated control filters with a size of $[B, N]$, where N \\ represents the length of each control filter. $\mathbf{Dis}$ and $\mathbf{Fx}$: the \\ disturbance and the filtered reference signal. Both of them \\ are of size $[B, F_s]$, where $F_s$ is the sampling rate.\\
\multicolumn{1}{l}{\rule{8.5cm}{0.2pt}}\\
$\textbf{Loss\_Calculation}(\mathbf{W}, \mathbf{Dis}, \mathbf{Fx}, B, N)$: \\
$\quad \mathbf{Pad} = \text{zeros}(B, N-1)$ ~~~$\triangleright$ Create a zero matrix. \\
$\quad \mathbf{Fx} = \text{concat}(\mathbf{Pad}, \mathbf{Fx}, \text{axis}=1) $ ~$\triangleright$ Concatenate matrices. \\
$\quad \mathbf{Fx\_v} = \text{unfold}(\mathbf{Fx}, \text{window\_size}=N, \text{stride}=1) $ ~$\triangleright$ Unfold \\ \quad with a window size of $N$ and a stride of 1. \\
$\quad \mathbf{Fx\_v} = \text{flip}(\mathbf{Fx\_v}, \text{dim}=3)$ ~$\triangleright$ Flip along the last dimension. \\
$\quad \mathbf{Y\_{\text{anti}}} = \mathbf{W} \times \mathbf{Fx\_v}^T $ ~~~$\triangleright$ Perform a matrix multiplication \\ \quad to get the anti-noise for suppressing the disturbance. \\
$\quad \mathbf{Err} = \mathbf{Dis} - \mathbf{Y\_{\text{anti}}} $ ~~~$\triangleright$ Compute the error signal. \\
$\quad \mathbf{Loss} = \text{mean}((\mathbf{Err})^2) $ ~$\triangleright$ The loss used for training CNN. \\
$\quad \textbf{Return Loss} $ \\
\multicolumn{1}{l}{\rule{8.5cm}{0.2pt}}\\
\end{tabular}
\label{Table 1}\vspace*{-0.1cm}
\end{table}

Compared to supervised learning, the proposed unsupervised approach can mitigate certain biases associated with the labelling process, including those induced by hyperparameter configurations. Additionally, unsupervised learning significantly reduces training costs by eliminating the labelling process. Furthermore, the unsupervised learning paradigm can quickly adapt to different tasks without being limited by labelled data, making it suitable for various ANC scenarios.

\begin{figure}[tp]
\centering
\subfigure{
\includegraphics[width=0.455\linewidth, height=3cm]{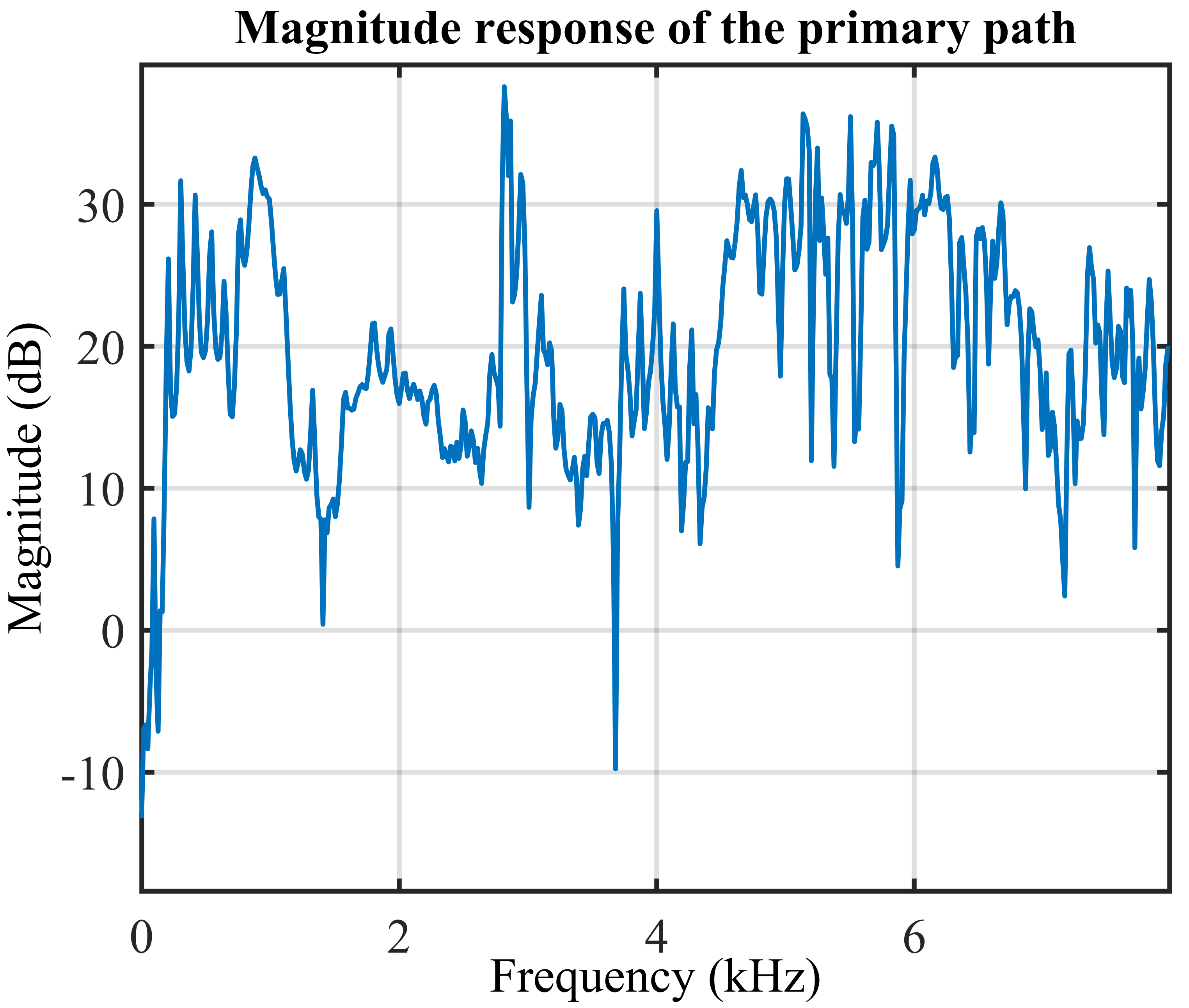}
\includegraphics[width=0.455\linewidth, height=3cm]{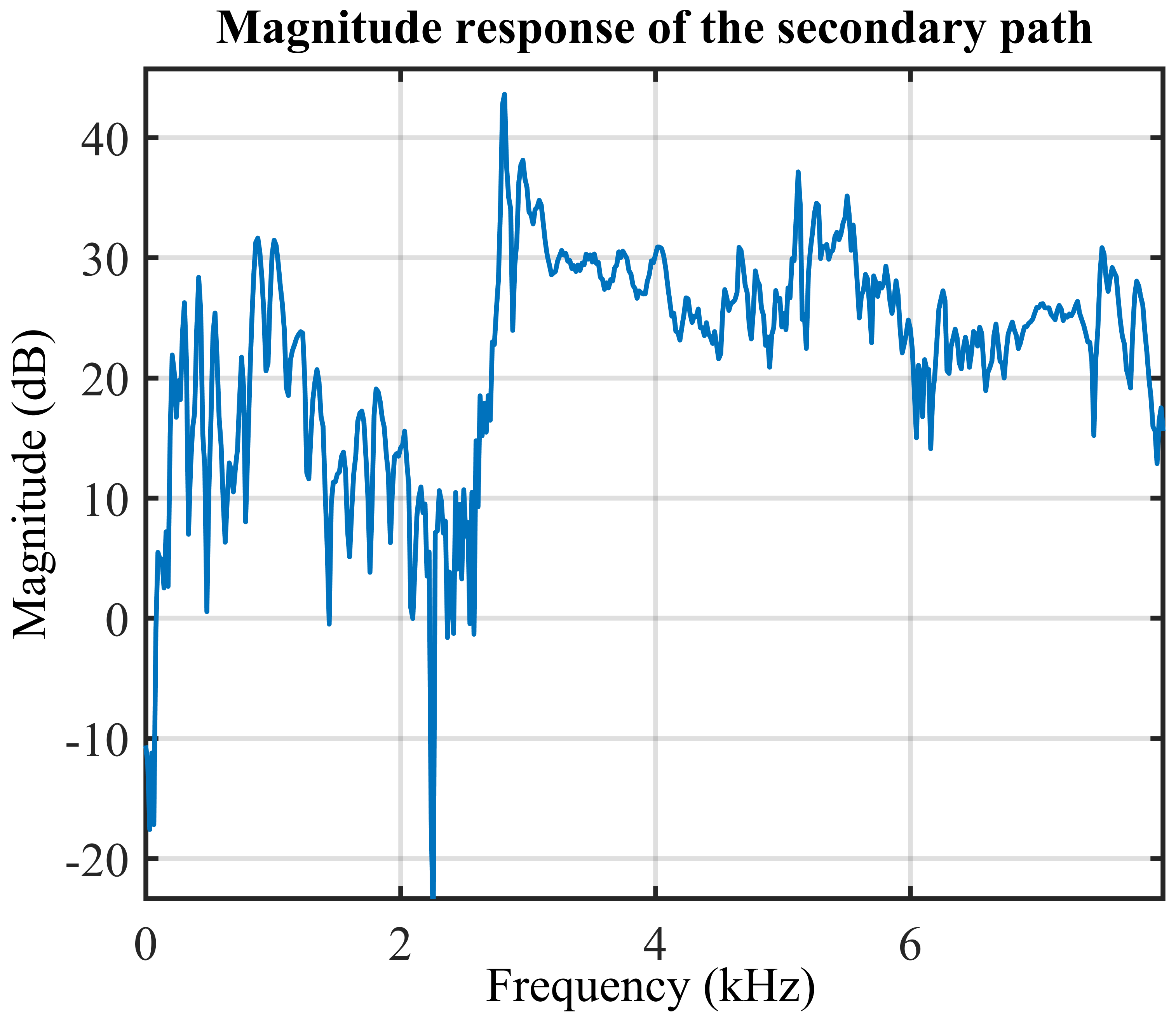}
}\vspace*{-0.3cm}
\subfigure{
\includegraphics[width=0.455\linewidth, height=3cm]{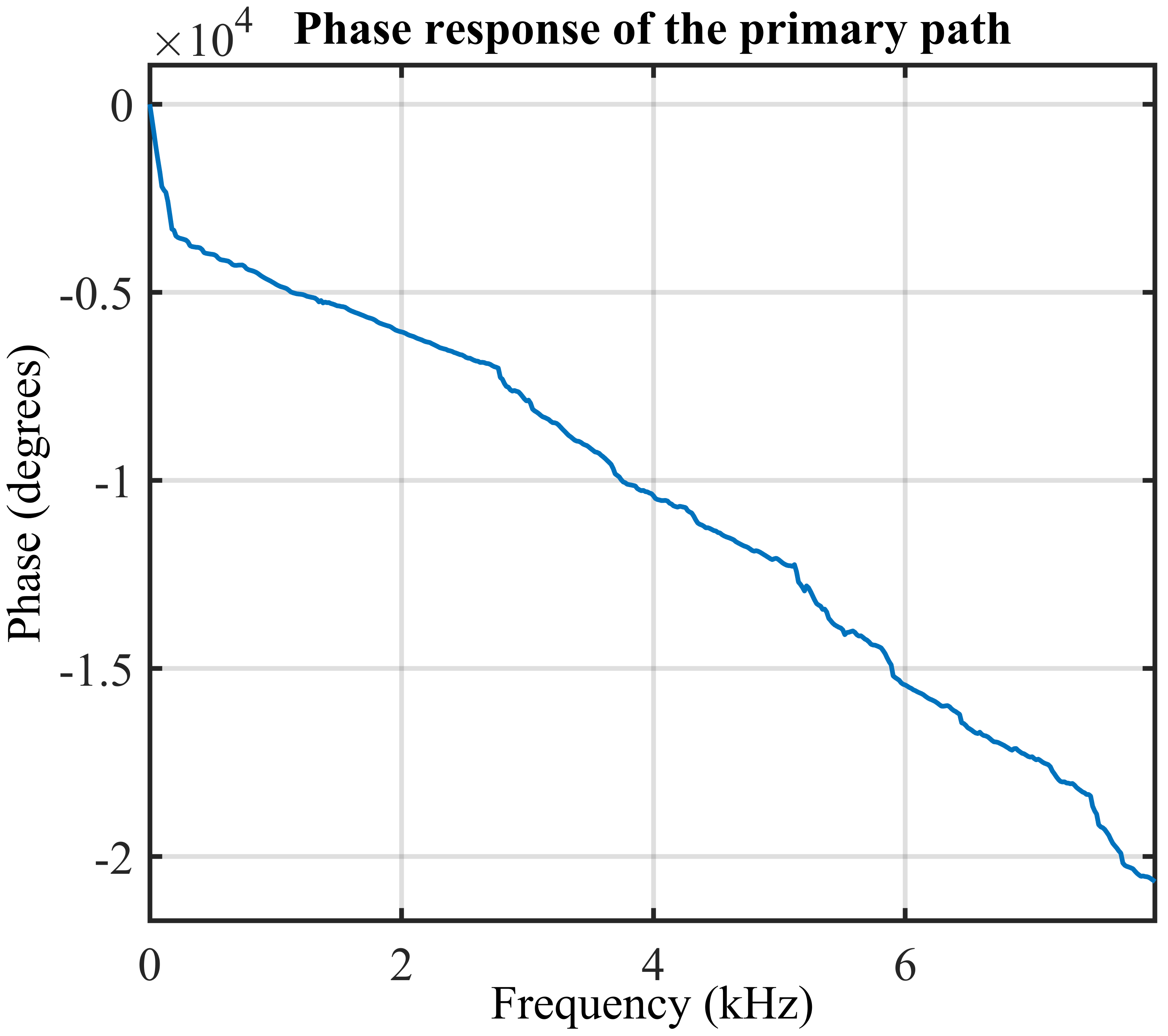}
\includegraphics[width=0.455\linewidth, height=3cm]{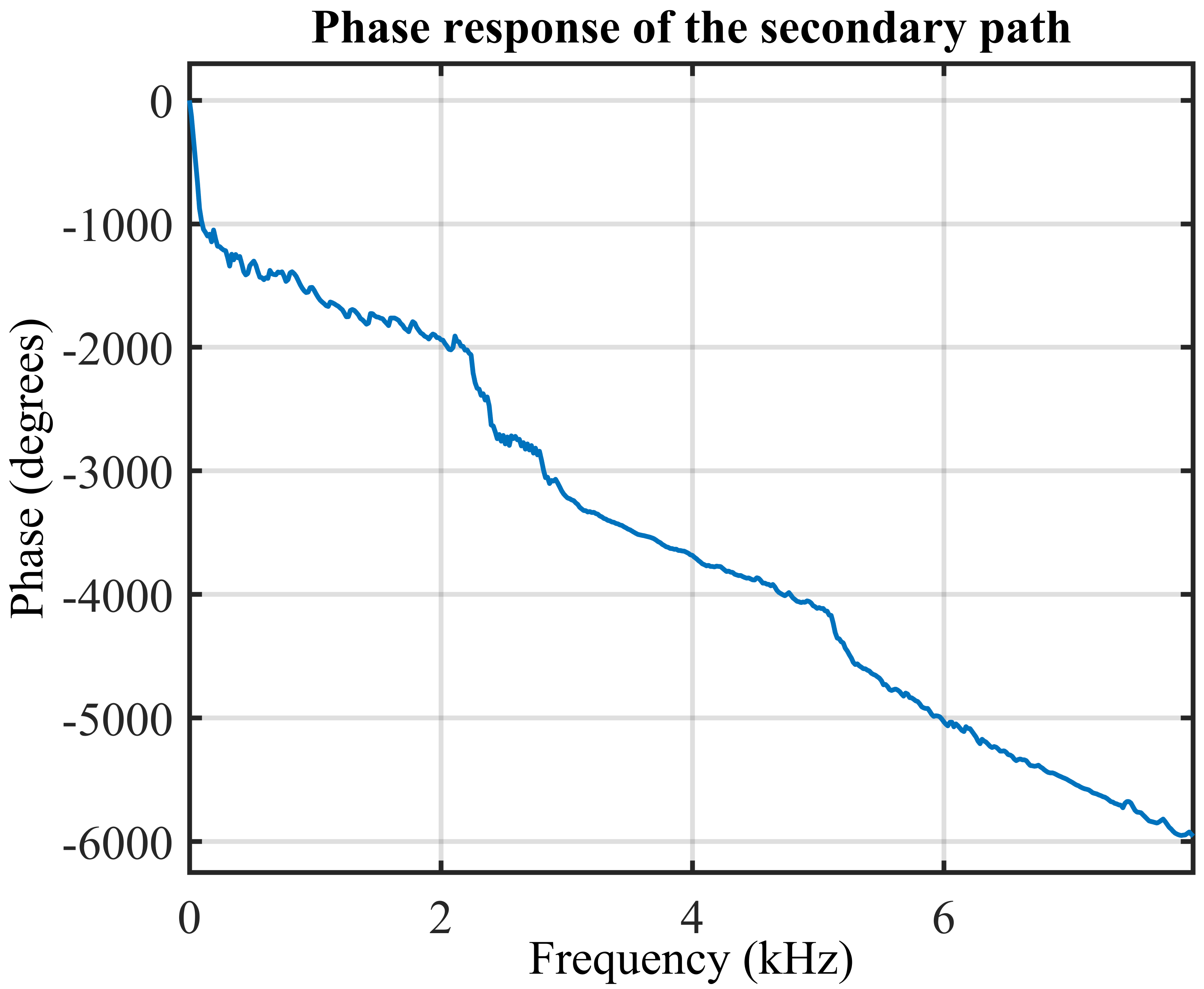}
}\vspace*{-0.3cm}
\caption{The magnitude responses and phase responses of the measured primary and secondary paths.}
\label{Fig 4}\vspace*{-0.3cm}
\end{figure}

\vspace*{-0.8cm}
\section{Numerical Simulations}
\vspace*{-0.2cm}
In terms of experimental setup, the number of sub control filters $M$ is $15$. The sampling rate $F_s$ and control filter's length $N$ are set to $16$ kHz and $1,024$ taps, respectively. A synthetic noise dataset containing $80,000$ noise instances is used to train the 1D CNN without additional labels. The noise instances are created by filtering white noise through bandpass filters with randomly chosen center frequencies and bandwidths. Each noise instance has a $1$-second duration. Synthetic bandpass filters are used as the primary path and secondary path during training the 1D CNN.

\begin{table}[tp]
\caption{NMSE (in dB) results for different ANC methods}\vspace*{0.1cm}
\centering
\begin{tabular}{c|c|c}
\hline
ANC algorithms & Aircraft noise & Drill noise\\
\hline
Unsupervised-GFANC & \textbf{-12.65} & \textbf{-9.04} \\
Supervised GFANC & -11.68 & -7.97 \\
FxLMS algorithm & -3.54 & -4.32 \\
\hline
\end{tabular}
\label{Table 2}\vspace*{-0.3cm}
\end{table}

In the following evaluations, the acoustic paths measured from a noise chamber's vent are used, as illustrated in Fig.~\ref{Fig 4}. The unsupervised-GFANC trained on synthetic acoustic paths is transferred to real acoustic paths. On the measured acoustic paths, sub control filters are obtained by decomposing the corresponding pre-trained broadband control filter, while keeping the trained 1D CNN model unchanged.

\begin{figure}[tp]
\centering
\subfigure{
\includegraphics[width=0.455\linewidth]{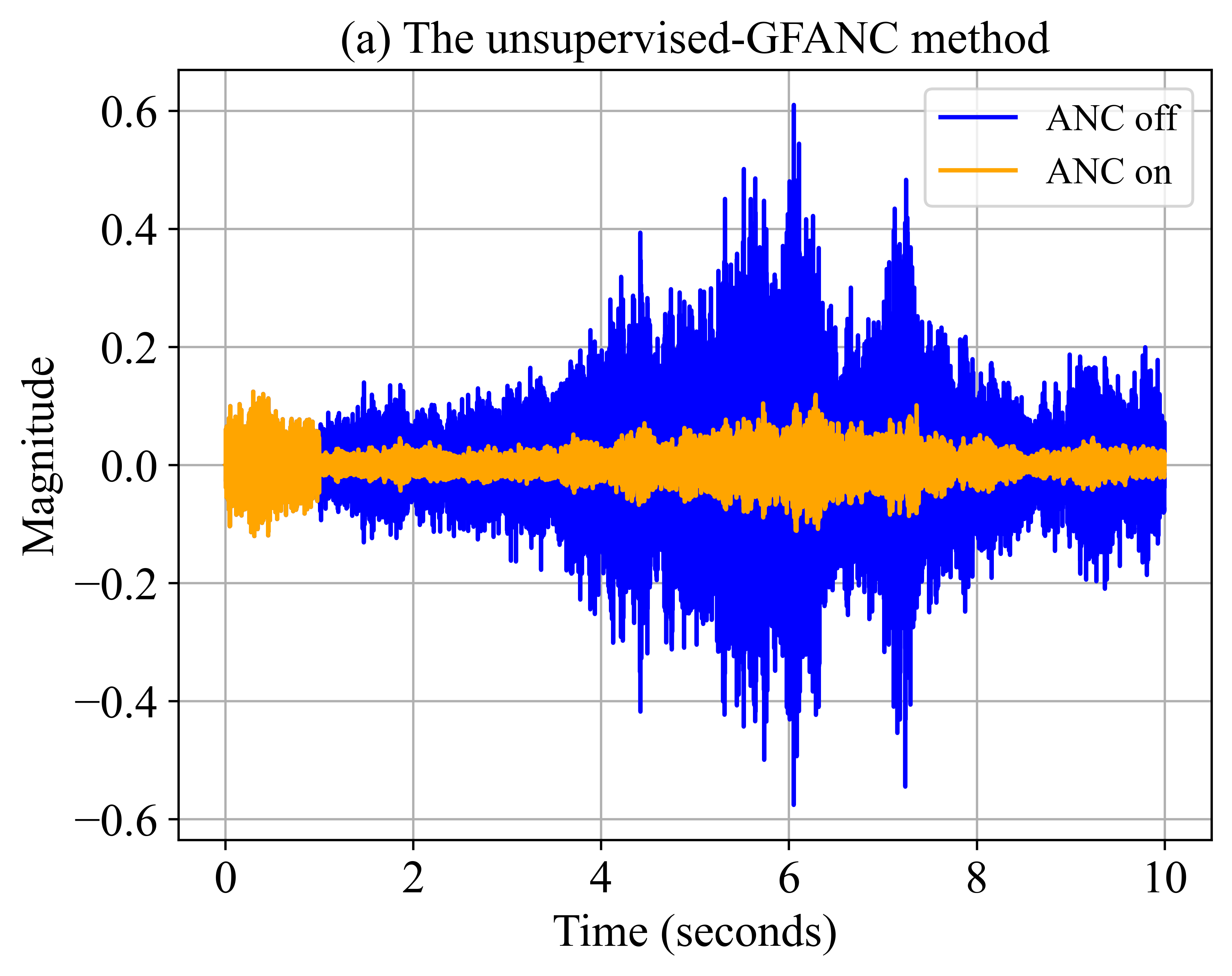}
}
\subfigure{
\includegraphics[width=0.455\linewidth]{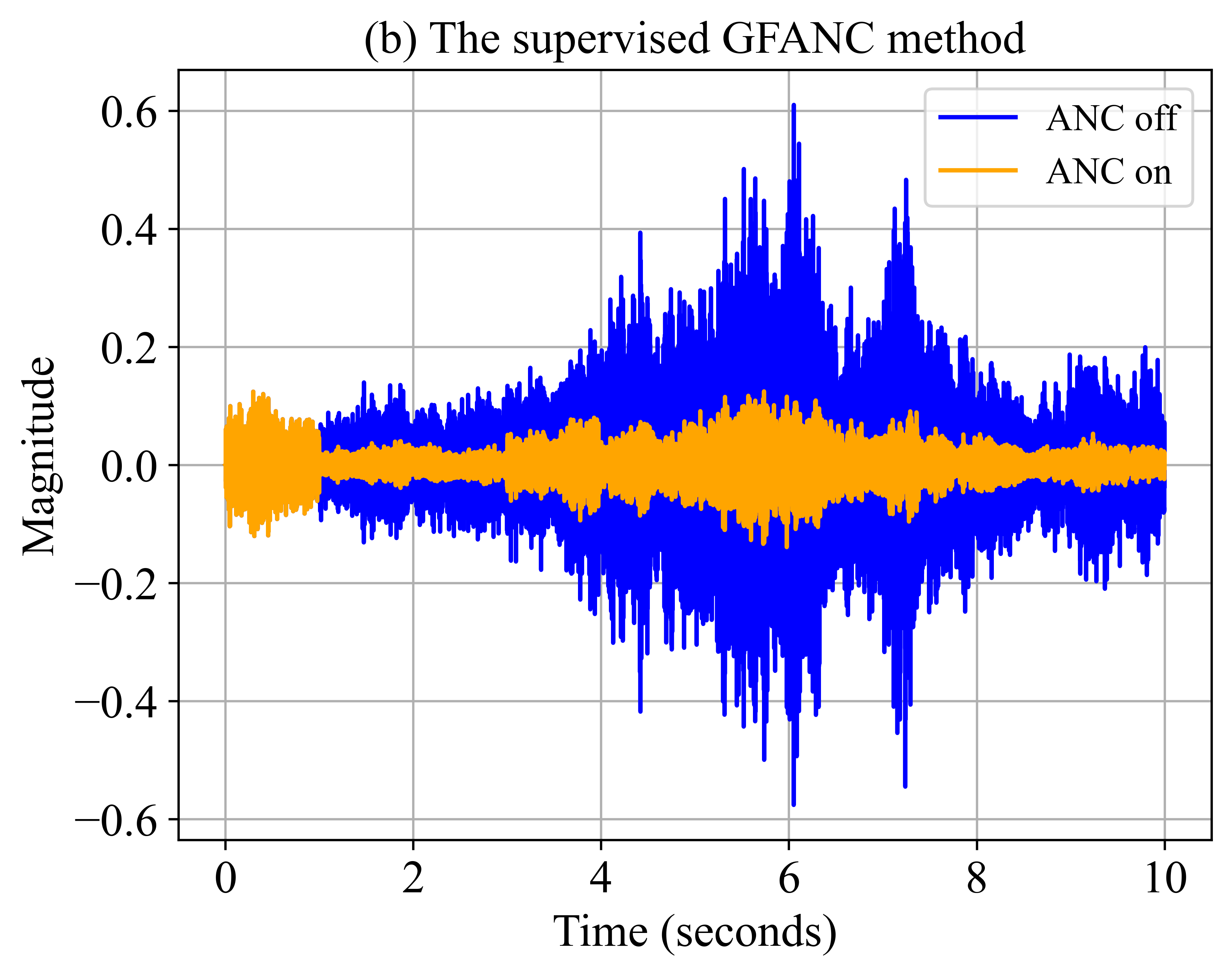}
}\vspace*{-0.3cm}
\subfigure{
\includegraphics[width=0.455\linewidth]{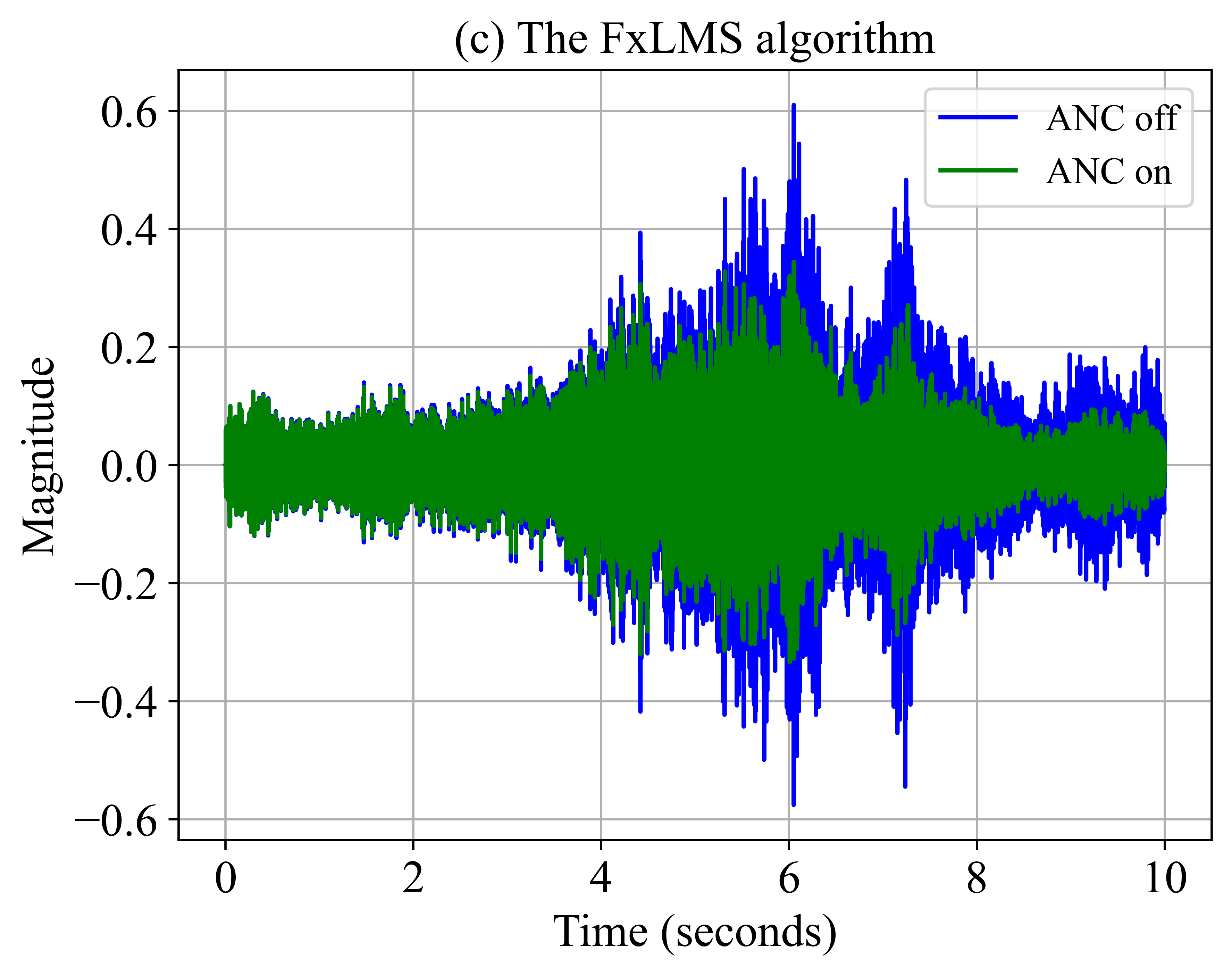}
}
\subfigure{
\includegraphics[width=0.44\linewidth]{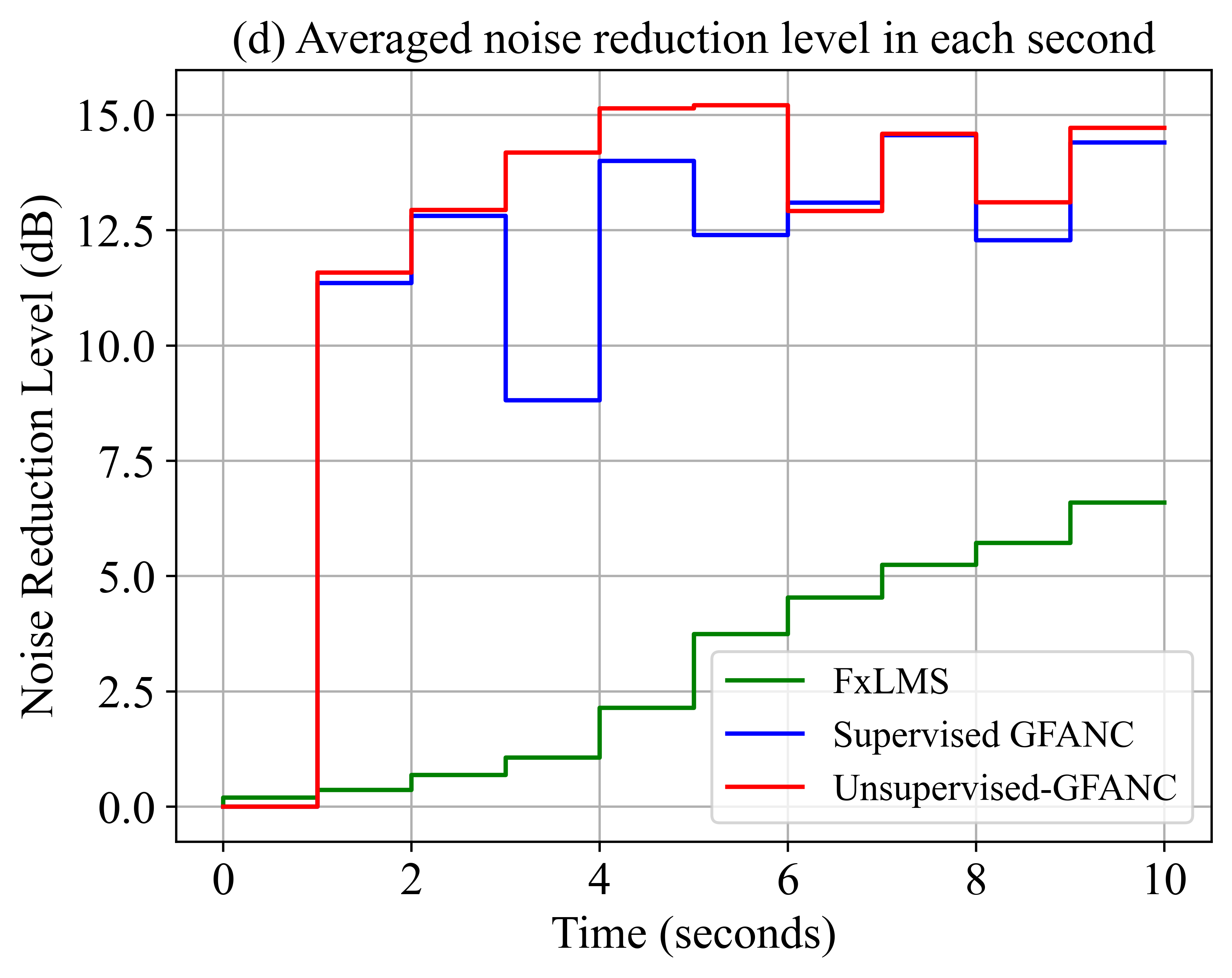}
}\vspace*{-0.3cm}
\caption{(a)-(c): Noise reduction results of different ANC algorithms, (d): Averaged noise reduction level in each second, on the aircraft noise.}\vspace*{-0.2cm}
\label{Fig 5}
\end{figure}

\vspace*{-0.3cm}
\subsection{Real-world noise cancellation on measured paths}
Firstly, the unsupervised-GFANC method is compared to the supervised GFANC method \cite{17} and the FxLMS algorithm with a step size of $0.0001$, in terms of normalized mean squared error (NMSE) in dB:
\begin{equation}
\mathrm{NMSE}=10 \log _{10} \frac{\sum_{n=1}^L e^2(n)}{\sum_{n=1}^L d^2(n)},
\end{equation}
where $L$ denotes the length of the signal vector. Typically, the value of NMSE is negative, and a lower value indicates superior performance. The NMSE comparison results on the aircraft and drill noises are summarized in Table \ref{Table 2}. The table shows that the unsupervised-GFANC method achieves superior NMSE results on both noise types. It is observed that the unsupervised-GFANC performs better than the supervised GFANC, possibly due to its capacity to mitigate certain biases associated with labelling procedures. Also, the NMSE obtained by the unsupervised-GFANC method exhibits a significant improvement over that obtained by the FxLMS algorithm. Therefore, this numerical simulation confirms the effectiveness of the unsupervised-GFANC method in the presence of real-world noises.

\begin{figure}[tp]
\centering
\subfigure{
\includegraphics[width=0.455\linewidth]{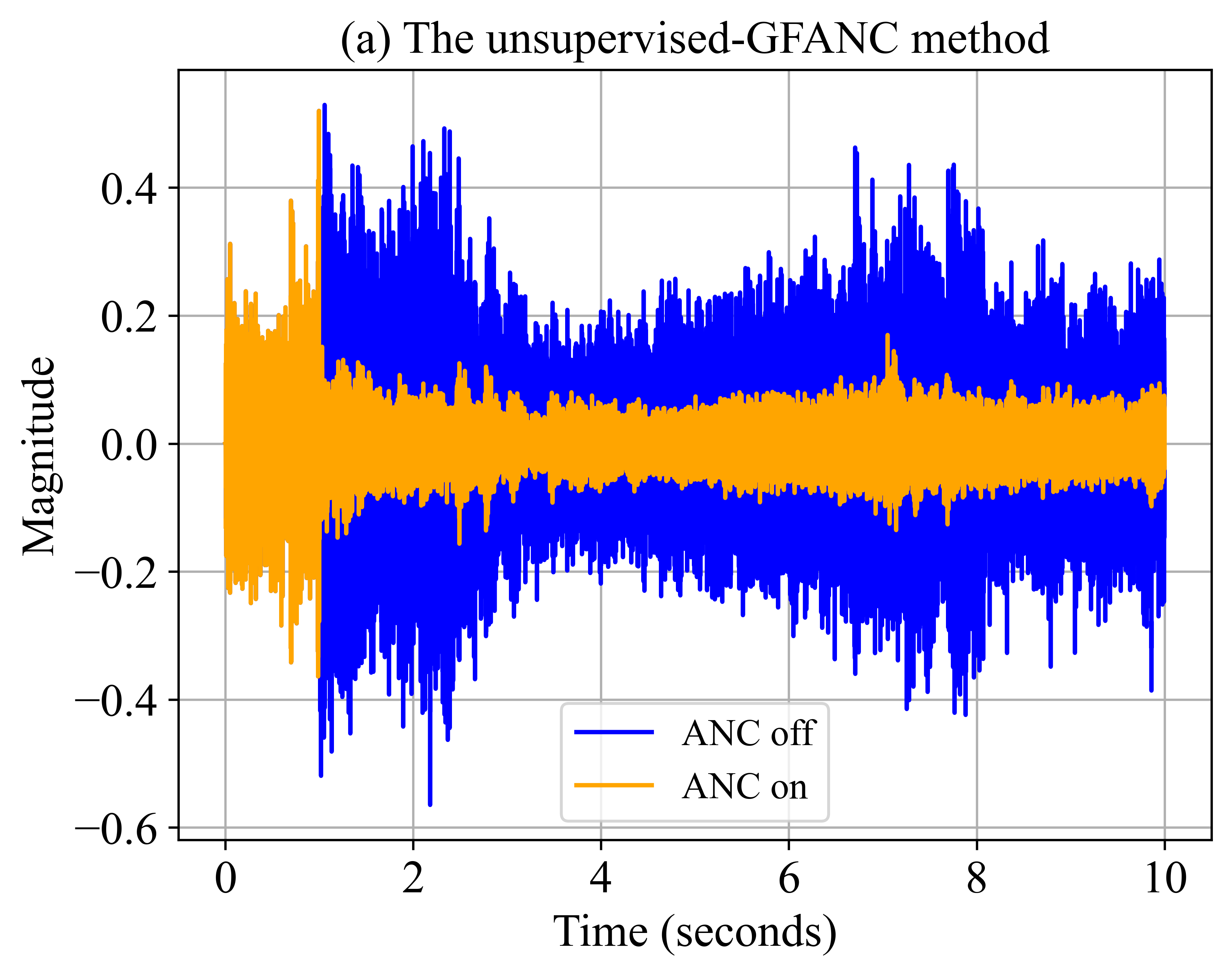}
}
\subfigure{
\includegraphics[width=0.455\linewidth]{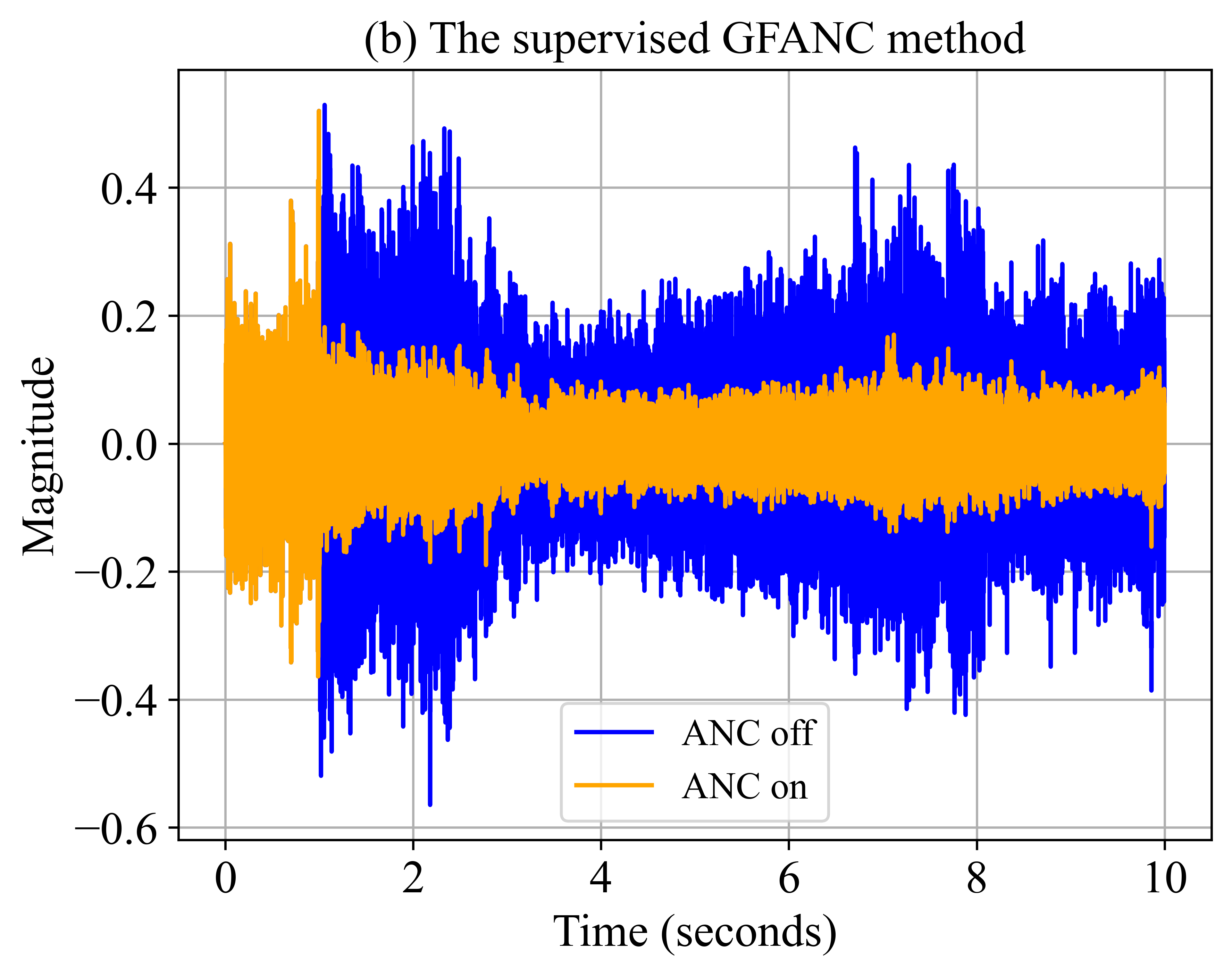}
}\vspace*{-0.3cm}
\subfigure{
\includegraphics[width=0.455\linewidth]{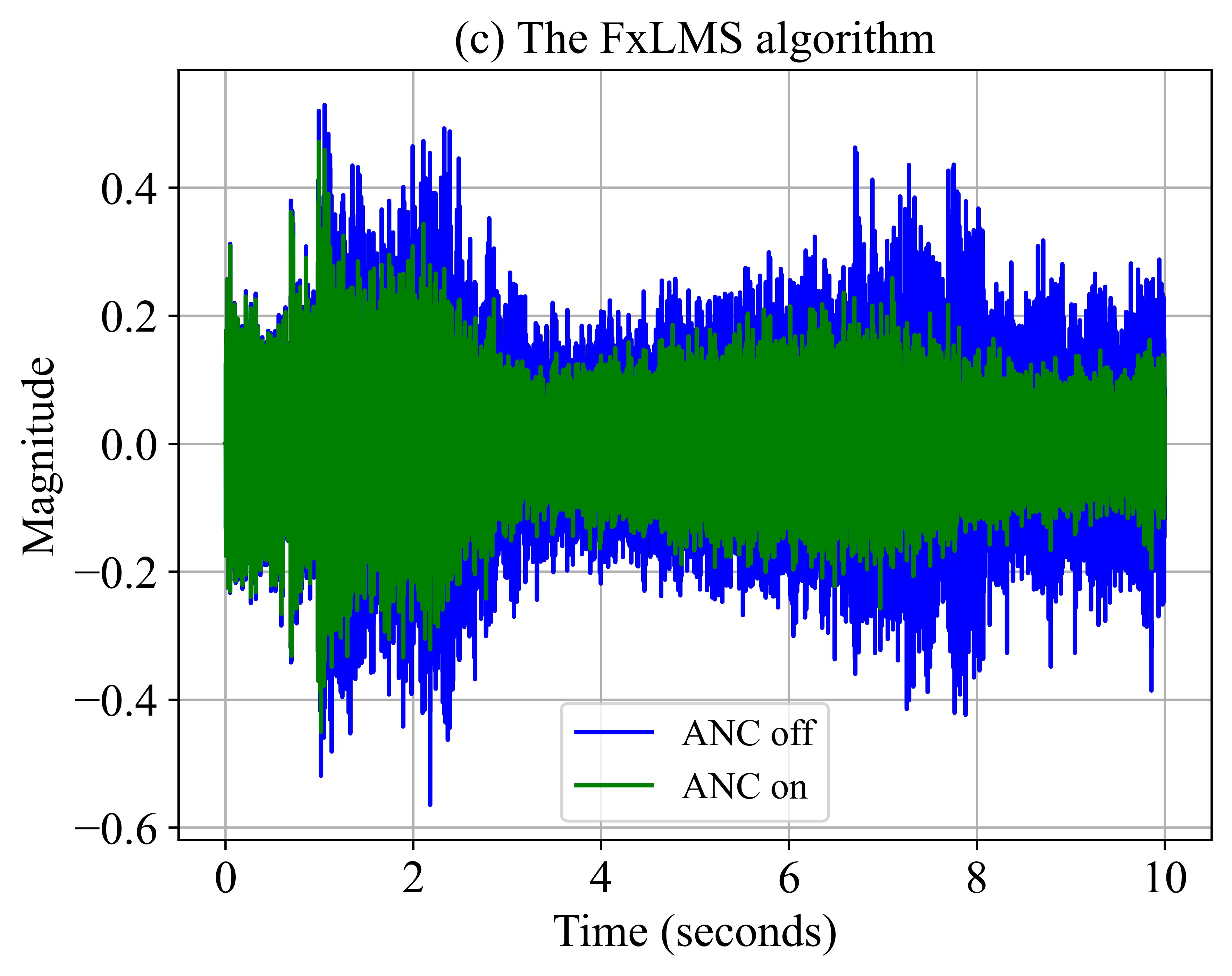}
}
\subfigure{
\includegraphics[width=0.44\linewidth]{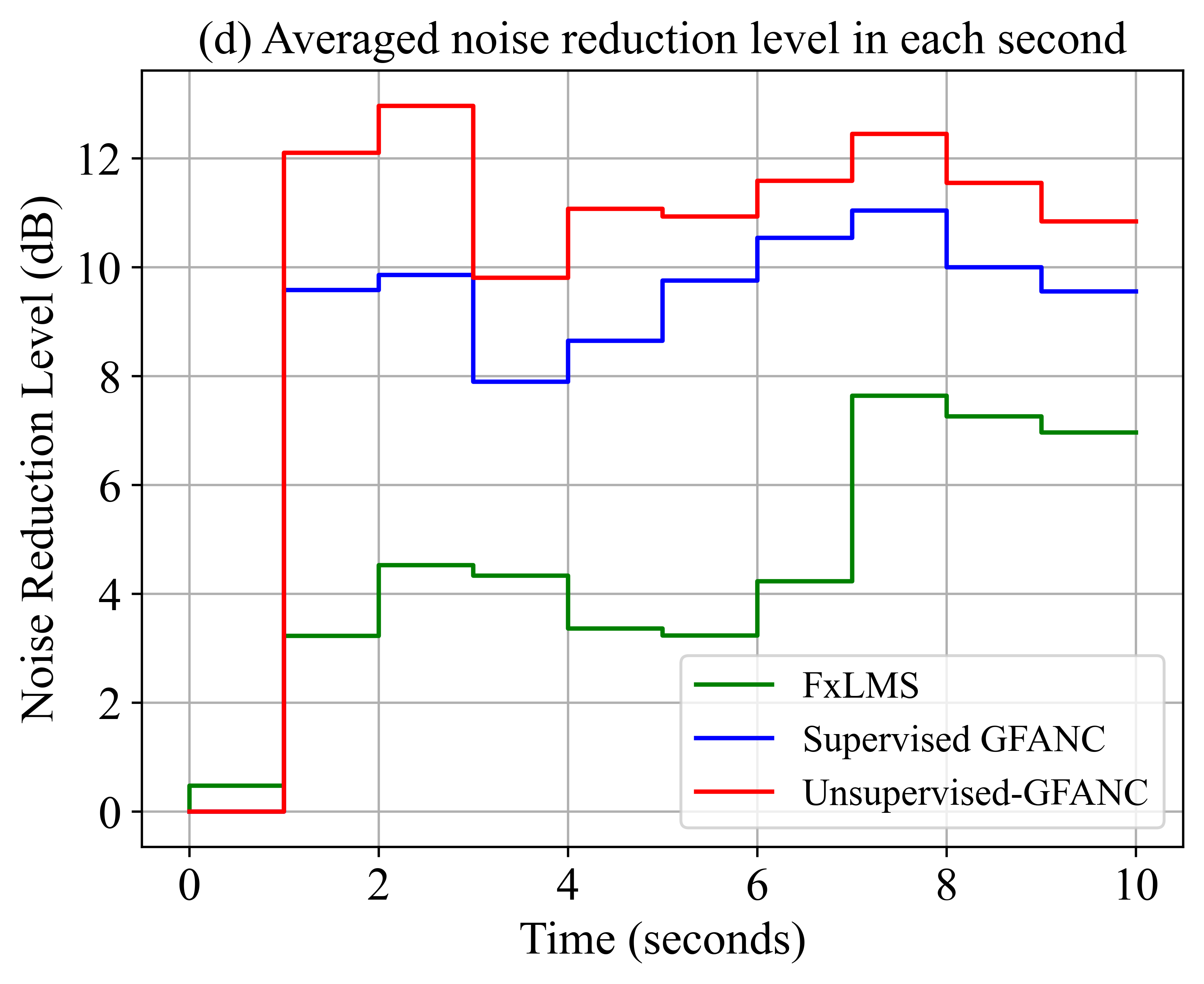}
}\vspace*{-0.3cm}
\caption{(a)-(c): Noise reduction results of different ANC algorithms, (d): Averaged noise reduction level in each second, on the drill noise.}\vspace*{-0.1cm}
\label{Fig 6}
\end{figure}

Fig.~\ref{Fig 5} and Fig.~\ref{Fig 6} show the noise reduction results using different ANC techniques on the aircraft and drill noises. Noticeably, the unsupervised-GFANC method performs much better than the FxLMS algorithm in terms of both response speed and noise reduction level. Also, the unsupervised-GFANC without labelling process achieves better performance than the supervised GFANC, enhancing its practicality for real-world applications. Taking the $3$s-$4$s aircraft noise (in Fig.~\ref{Fig 5}) as an example, the averaged noise reduction level of the unsupervised-GFANC is about $12$ dB and $5$ dB higher than that of FxLMS and supervised GFANC, respectively. However, the unsupervised-GFANC and supervised GFANC have no noise cancellation in the first second, because the initial control filter coefficients are set to zero \cite{22}.

\vspace*{-0.2cm}
\subsection{Analysis of Power Spectral Density}
\begin{figure}[tp]
\centering
\subfigure{
\includegraphics[width=0.455\linewidth]{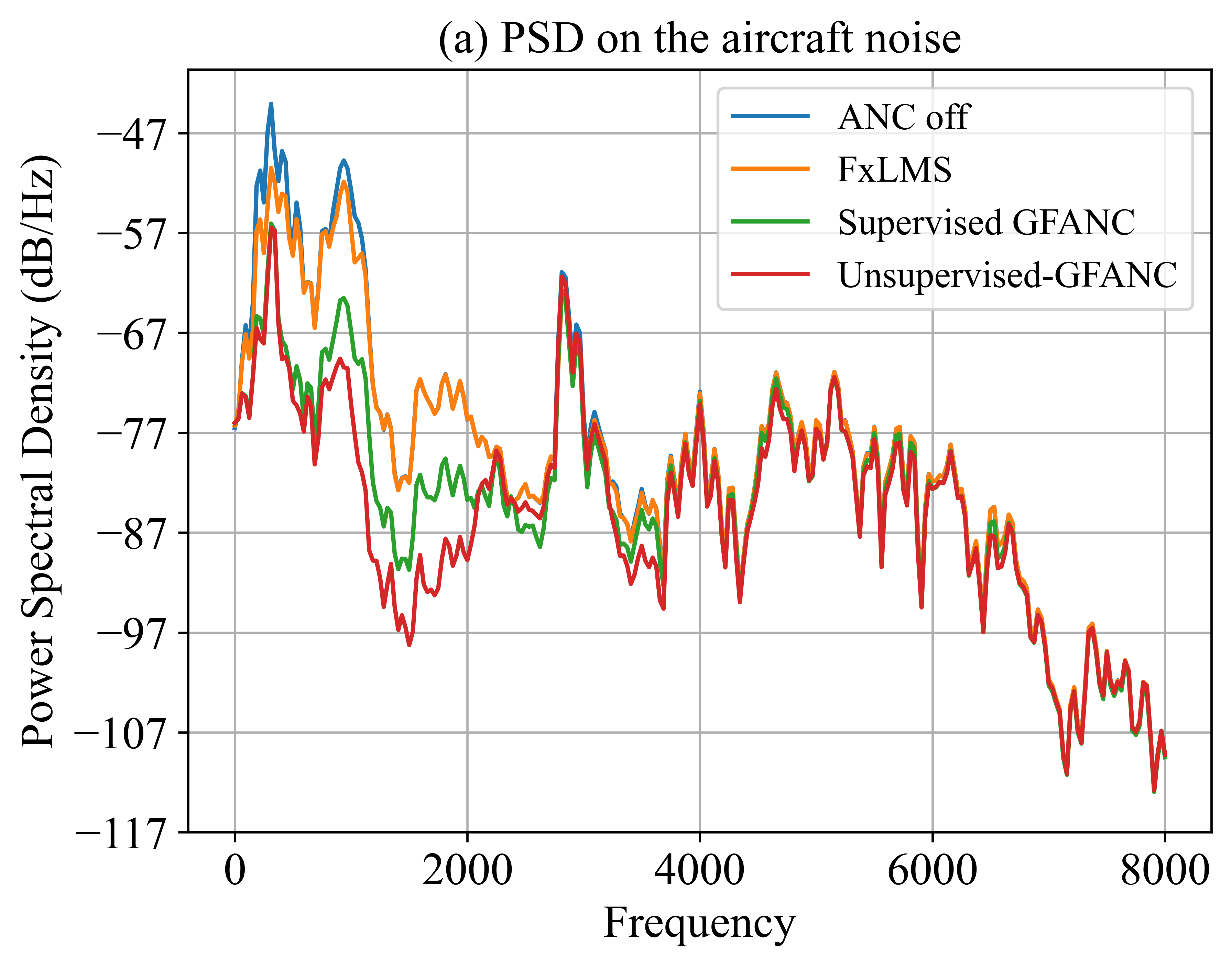}
}
\subfigure{
\includegraphics[width=0.455\linewidth]{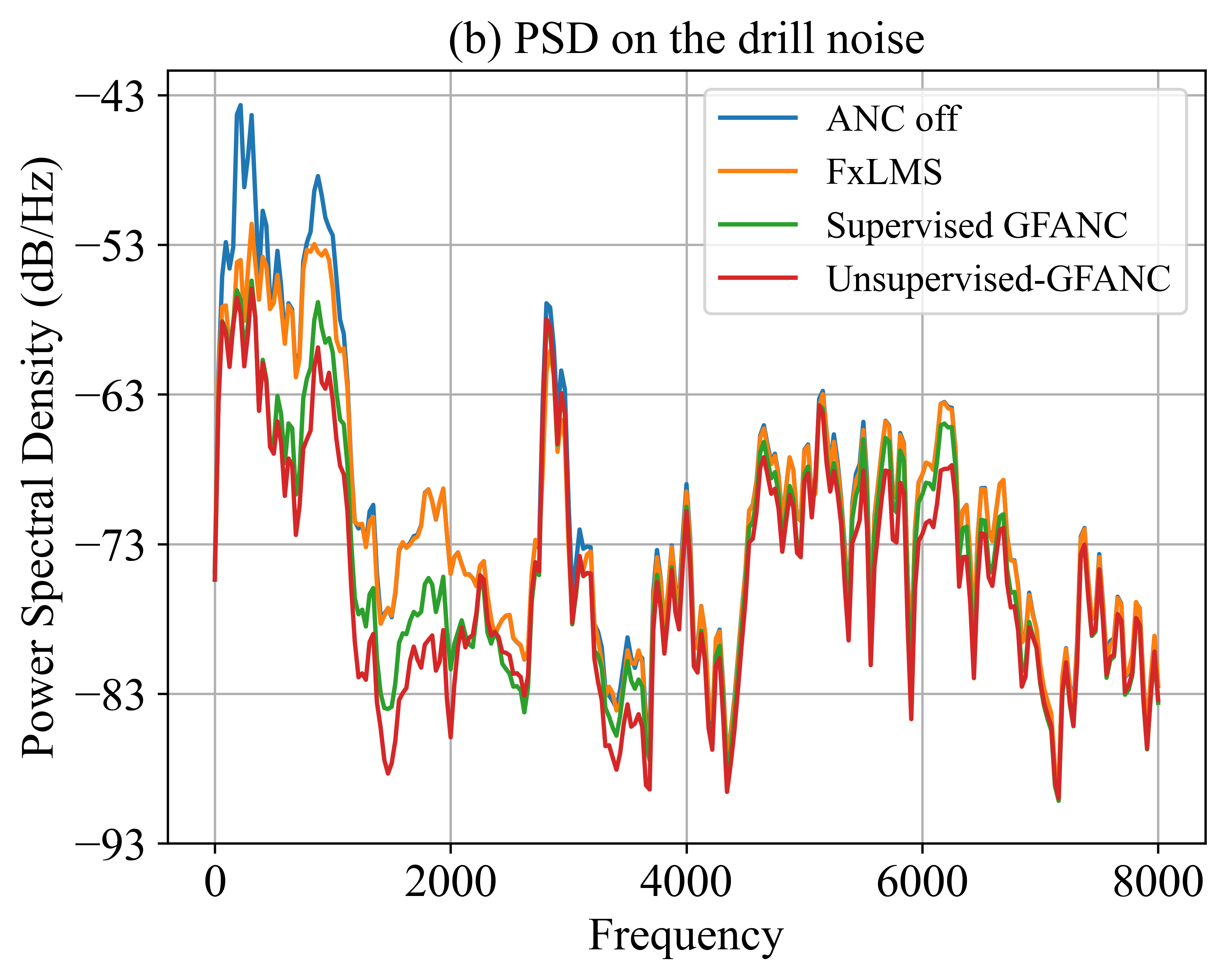}
}\vspace*{-0.3cm}
\caption{The power spectral density (PSD) of the aircraft noise (a) and drill noise (b) by different ANC algorithms.}
\label{Fig 7}
\end{figure}

Furthermore, the noise reduction performances of these ANC algorithms are assessed by analyzing the power spectral density (PSD) in Fig.~\ref{Fig 7}. PSD can provide insights into the distribution of power across different frequencies within the noise signal. Noticeably, the noise components below $2,000$ Hz are effectively attenuated by the unsupervised-GFANC and supervised GFANC. In comparison, the FxLMS algorithm is less effective at removing the two types of noise, probably due to its limited ability to track rapidly varying noises. Therefore, the effectiveness of the unsupervised-GFANC method in cancelling low-frequency noises is verified.

\vspace*{-0.1cm}
\section{Conclusion}
\vspace*{-0.1cm}
This paper proposes a novel unsupervised-GFANC method and outlines its ANC performance. During training the 1D CNN, the co-processor and real-time controller are integrated into an end-to-end derivable ANC system. With the backpropagation of derivatives, the accumulated squared error signal is used as a training loss to update the network parameters. Without labelling procedures, the proposed unsupervised-GFANC method can mitigate certain labelling biases and reduce annotation costs. The results of reducing real noises demonstrate the effectiveness of the unsupervised learning paradigm. Also, the proposed method trained on synthetic acoustic paths can be easily transferred to real acoustic paths, which indicates its good transferability and practicality. Furthermore, this unsupervised learning approach can be applied to other deep learning-based ANC methods.

\newpage
\bibliographystyle{IEEEbib}
\small
\bibliography{A}
\end{document}